\documentclass[%
 reprint,
 amsmath,amssymb,
 aps,
 onecolumn,
]{revtex4-2}

\usepackage{graphicx}
\usepackage{dcolumn}
\usepackage{bm}
\usepackage[mathlines]{lineno}

\begin{document}

\preprint{APS/123-QED}

\title{\textbf{prompt acceleration of a short-lifetime low-energy muon beam}}

\author{Xiao-Nan Wang}
 \altaffiliation[Also at ]{Physics and Space Science College, China West Normal University}
 \author{Xiao-Fei Lan}%
 \email{lan-x-f@163.com}
\affiliation{%
Physics and Space Science College, China West Normal University, Nanchong 637009, China
}%

\author{Yong-Sheng Huang}
\email{huangys82@ihep.ac.cn}
\affiliation{
 Institute of High Energy Physics, CAS, Beijing 100049, China
}%

\author{Hao Zhang}
\altaffiliation[Also at ]{Department of Physics, National University of Defense Technology}
\author{Tong-Pu Yu}
\affiliation{%
 Department of Physics, National University of Defense Technology, Changsha 410073, China
}%
\begin{abstract}
 An energetic muon beam is an attractive key to unlock new physics beyond the Standard Model: the lepton flavor violation or the anomalous magnetic moment, and also is a competitive candidate for the expected neutrino factory. Lots of the muon scientific applications are limited by low flux cosmic-ray muons, low energy muon sources or extremely expensive muon accelerators. An prompt acceleration of the low-energy muon beam is found in the beam-driven plasma wakefield up to $\mathrm{TV/m}$. The muon beam is accelerated from $275\mathrm{MeV}$ to more than $10\mathrm{GeV}$ within $22.5\mathrm{ps}$. Choosing the injection time of the muon beam in a proper range, the longitudinal spatial distribution and the energy distribution of the accelerated muon beam are compressed. The efficiency of the energy transfer from the driven electron beam to the muon beam can reach $20\%$. The prompt acceleration scheme is a promising avenue to bring the expected neutrino factory and the muon collider into reality and to catch new physics beyond the Standard Model.
\end{abstract}

\maketitle

An energetic muon beam is a potential option to explore new physics beyond the Standard Model: the lepton flavor violation\textsuperscript{\cite{flavorvio}}\textsuperscript{\cite{raredacay}}\textsuperscript{\cite{berger2014mu3e}} or the anomalous magnetic moment\textsuperscript{\cite{farley200447}}\textsuperscript{\cite{charpak1961measurement}}\textsuperscript{\cite{bailey1979final}}, and also is a competitive candidate for the expected neutrino factory\textsuperscript{\cite{neutrinofac}}\textsuperscript{\cite{cao2014muon}}. Besides, a muon source with some unique properties has widely applications, such as muon spin rotation, resonance and relaxation ($\mu$SR)\textsuperscript{\cite{borozdin2003radiographic}}\textsuperscript{\cite{alvarez1970search}}\textsuperscript{\cite{borozdin2012cosmic}}\textsuperscript{\cite{priedhorsky2003detection}}, the muon catalyzed fusion\textsuperscript{\cite{jones1986observation}}, non-destructive elemental depth-profiling with muonic X-rays\textsuperscript{\cite{kubo2008non}}, the future muon colliders\textsuperscript{\cite{ankenbrandt1999status}}, the measurement of the muon rare decay\textsuperscript{\cite{berger2014mu3e}}\textsuperscript{\cite{bartoszek2015mu2e}}\textsuperscript{\cite{kutschke2009mu2e}}\textsuperscript{\cite{grassi2005meg}}\textsuperscript{\cite{kuno2013search}}, the fine measurement of atomic structure\textsuperscript{\cite{pohl2010size}}\textsuperscript{\cite{antognini2013proton}}  and so on.

   However, until now, there are two types of muon sources: the high-energy low-flux cosmic muon source\textsuperscript{\cite{nagamine2003introductory}}\textsuperscript{\cite{bose1944cosmic}} and the low-energy  Conventional muon Sources\textsuperscript{\cite{fermi1950high}}\textsuperscript{\cite{tsai1974pair}}\textsuperscript{\cite{athar2001muon}}. The low-energy muon has short lifetime and the cosmic muon source has so small flux to be used for the study of new physics, such as  rare decay measurements or anomalous magnetic moments. To build a muon collider is also an extremely expansive expectation\textsuperscript{\cite{carne1991isis}}\textsuperscript{\cite{miyake2009j}}\textsuperscript{\cite{abela1994musr}}\textsuperscript{\cite{marshall1992muon}}.

Recently, the laser-plasma wakefield breaks the limit of the traditional accelerator and can reach several GV/m to hundreds of GV/m\textsuperscript{\cite{leemans2014multi}}\textsuperscript{\cite{chen1985acceleration}}\textsuperscript{\cite{litos2014high}}. Some proposals on the muon beam acceleration in the laser-plasma wakefield have been given\textsuperscript{\cite{Zhang2018All}}. There is a muon trapping energy threshold of about GeV to ensure that the muon can be captured by the laser wakefield. However, how to accelerate the low-energy muon from two hundreds of MeV to several GeV quickly is the key challenge to obtain the high-energy muon beam.

  In the linear regime, the maximal acceleration gradient excited by an electron beam is proportional to $\frac{N}{\sigma_x^2}$\textsuperscript{\cite{sahai2019schemes}}, where N is the number of the driven electron beam, $\sigma_z$ is the bunch length. To reach higher acceleration gradient, the electron bunch must have larger charge number and shorter beam length. A three GeV electron bunch with the charge of 16.7 nanocoulomb and the beam length of 100 femtoseconds, can be obtained from the laser wakefield acceleration\textsuperscript{\cite{hogan2010plasma}}. With this electron beam, the driven plasma wakefield can reach $\mathrm{TV/m}$, which is expected to conquer the challenge of the low-initial-energy muon acceleration.

\begin{figure*}
\includegraphics[scale=0.1]{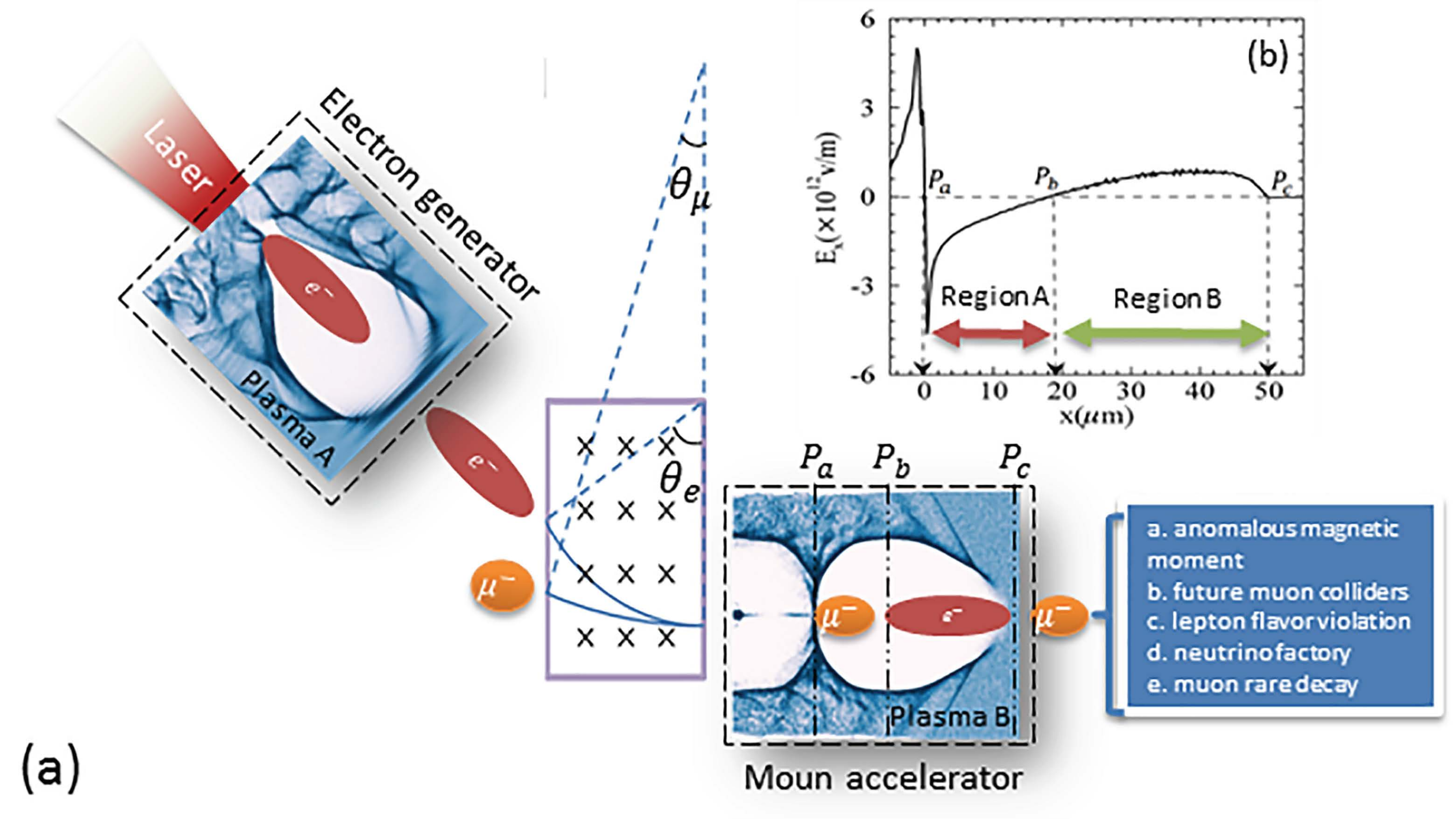}
\caption{\label{fig1} \small \textbf{The scheme for the prompt acceleration of the low-initial-energy muon beam. The typical longitudinal wakefield $E_x$}. (a) shows the scheme for the prompt acceleration. The laser is injected into the electron generator. Since the laser has the normalized field $a_0=53$, the spot size $\omega_0=10\mathrm{\mu m}$ and the pulse width $\tau=33\mathrm{fs}$, an electron beam with the energy of 3GeV and the charge of 16.7 nanocolomb can be generated at the exit point of the plasma A. The electron beam is injected into the plasma B through a magnetic field. In plasma C, a plasma wakefield of about several $\mathrm{TV/m}$ can be excited by the electron beam. The muon beam is also deflected by the same magnetic field and then is injected into the bubble field in plasma B. The position $P_a$, $P_b$, $P_c$ on the muon accelerator corresponds to the tree zero points in Figure 1 (b), respectively. (b) shows the typical distribution of longitudinal field $E_x$ on the center line of the blow-out regime at $t=0.225\mathrm{ps}$ after the injection of the electron bunch. The electron beam is located in region B close to the point $P_c$ and loses energy. For muons, the point, $P_b$ is the critical point between the acceleration region, region A, and the deceleration region, region B. If the injection point of the muon beam is near $P_b$, the longitudinal spatial distribution of the muon beam is compressed and also the energy spread is decreased. There is an optimal injection point, i.e., an optimal injection time.}
\end{figure*}

 Figure \ref{fig1} (a) shows a compact scheme for the prompt acceleration of the low-initial-energy muon beam using the 'blowout regime' of plasma wakefield\textsuperscript{\cite{lu2006nonlinear}}. This scheme has two parts: an electron generator and a muon accelerator. In the first part, the laser beam injects into the plasma A and excit the plasma wakefield due to the pondermotive force of the laser. An electron beam with the energy of 3GeV and the charge of 16.7 nanocolomb is produced in this wakefield. The electron beam is deflected in a dipole magnetic field and is injected into plasma B sequently. In the second part, the muon beam is also defleted an angle, $\theta_{\mu}$ by the dipole magnetic field. At the end of the magnetic field, the electron beam B and the muon beam are injected into plasma C sequently. In plasma C, the plasma wakefield of several $\mathrm{TV/m}$ is excited. The first bubble driven by the electron bunch moves forward with the group velocity closing the speed of light in a vacuum. The followed muon beam is also located in the wakefield of the first bubble. As an example, the key points of the prompt acceleration are shown under the parameters:
  $n_p=1\times10^{19}\mathrm{cm^{-3}}$, $n_b=7n_p=7\times10^{19}\mathrm{cm^{-3}}$, $\sigma_x\approx\frac{6.3c}{\omega_p}$ and $\sigma_y=\sigma_z\approx\frac{2.1c}{\omega_p}$, $E_e\approx3\mathrm{GeV}$, where $n_p$ is the plasma density, $n_b$ is the maximum of the electron bunch density $\sigma_x,\sigma_y=\sigma_z$ is the transverse and the longitudinal size of the electron beam, respectively, $\omega_p$ is the plasma frequency, $E_e$ is the eletron energy.

  We focus on the prompt acceleration of a muon beam in the beam-driven plasma wakefield as follows. The typical distribution of the longitudinal wakefield $E_x$ is shown in Figure \ref{fig1} (c). $P_a$, $P_b$, $P_c$  are located at the zero point of $E_x (x)$. $P_b$ is the demarcation point between region A and region B. In region B, the field is positive and can decelerate muons. In region A, the field is negative and can accelerate muons. The maximum acceleration field is up to $5.27\mathrm{TV/m}$. The muon's energy gain can be estimated by the integration of the acceleration field over the muon acceleration length. Located in the maximum acceleration field, a muon can obtain energy of about 1$\mathrm{GeV}$ per picosecond. In the beam-driven plasma wakefield, the final acceleration energy of muons mainly depends on the initial energy and the injection time. The influence relationships will be shown in the following discussions. When the position is closer to point b, the absolute values of the electric field in region A and B are both decreased slowly to zero as shown in Figure \ref{fig1} (c). In the non-uniform longitudinal wakefield $E_x$, the longitudinal size of the muon beam will be compressed. Due to this compression, the muon beam also has a small final energy spread. There is an optimal injection time for the energy spread. The energy spread of the accelerated muon beam can reach about $6\%$.

\begin{figure*}
\includegraphics[scale=0.1]{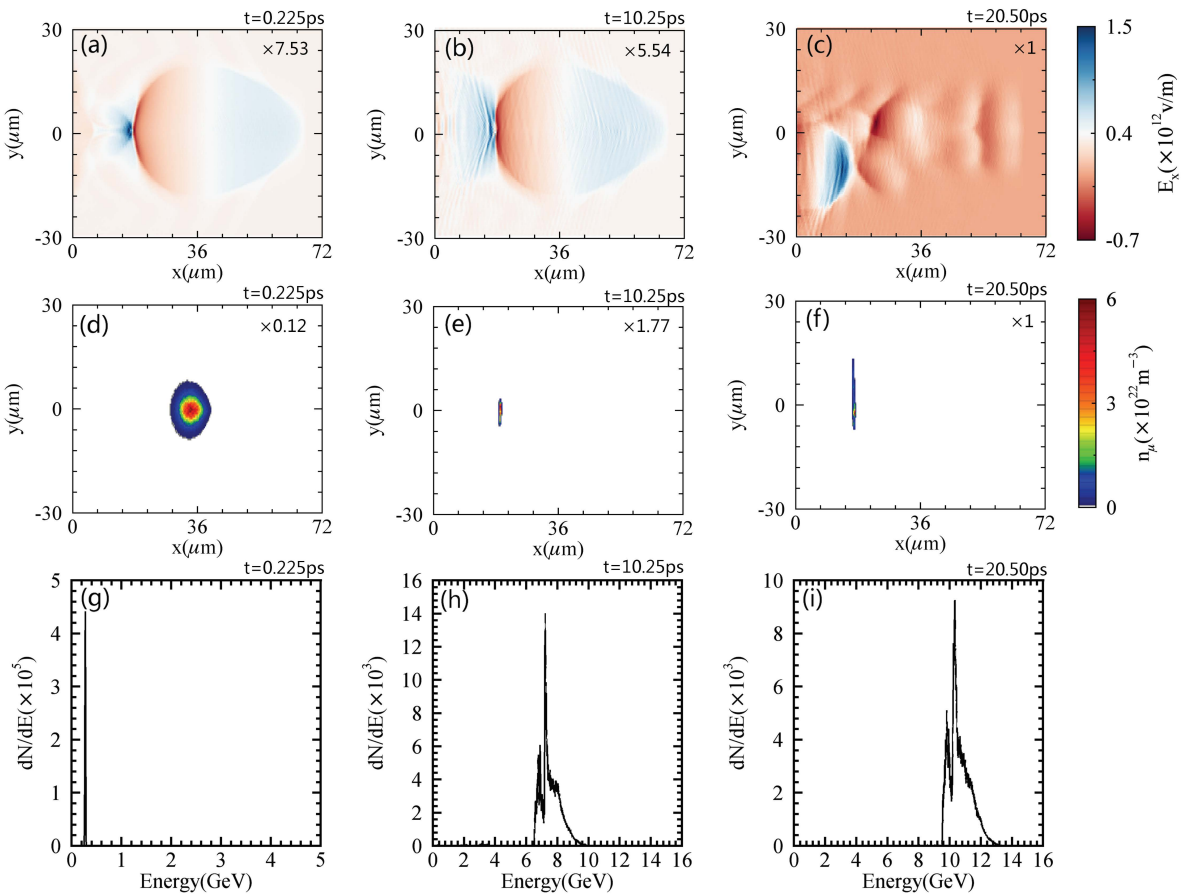}
\caption{\label{fig2} {\small \textbf{The evolution of the beam-driven plasma wakefield, the muon density, and the muon energy spectrum}. (a-i) the snapshots are taken at the simulation time $t=0.225\mathrm{ps}$, $t=10.25\mathrm{ps}$, $t=22.50\mathrm{ps}$ respectively. (a-c) is the longitudinal field $E_x$ at the plane of $z=0$. (d-f) is the evolution of space distribution of muon density. (g-i) is the energy spectrum of the muon beam at the corresponding moment.}}
\end{figure*}

 The three-dimensional Particle-in-cell (PIC) simulation results show the evolution of the muon acceleration process in the beam-driven plasma wakefield in detail. The parameters of the simulation are listed in method. Figure  \ref{fig2} (a) shows that the wakefield is formed. After that, the structure of this wakefield remains stable and the maximum acceleration field keeps more than $\mathrm{TV/m}$ shown in Figure \ref{fig2} (b). Figure  \ref{fig2} (c) shows that the energy of the electron beam was depleted, leading to the break of the wakefield. As a result, the muon acceleration process in this wakefield is completed.

  Figure \ref{fig2} (a,d) show that the muon beam locates at the front part of acceleration field. The velocity of the muon beam is smaller than that of the wakefield. Therefore, it goes toward the rear edge of the wakefield shown in Figure \ref{fig2} (b,e). Figure \ref{fig2} (d,e) shows that the longitudinal size of the muon beam is compressed from $6.66\mathrm{\mu m}$ to about $2\mathrm{\mu m}$ within a few picoseconds. The energy gain of the muon located at the front of the muon beam is lower than those located at the tail of the muon beam. Therefore the muon located at the front of the muon beam will be closer and closer to those located at the tail of the muon beam. The muon beam would be compressed in longitudinal direction. Figure \ref{fig2} (i) shows that the peak energy is up to more than 10GeV. The energy spread $\frac{\triangle E}{E}$ is about $6\%$ after the acceleration in the non-uniform longitudinal wakefield $E_x$.

  In the laser wakefield acceleration\textsuperscript{\cite{Tajima1979Laser}}\textsuperscript{\cite{wang2013quasi}}\textsuperscript{\cite{martins2010exploring}}, there is a muon trapping energy threhold\textsuperscript{\cite{Zhang2018All}} to ensure that the muon can be captured by the wakefield. If the initial energy of the muon is equal to the trapping energy threshold, the muon's velocity must equal to that of the wakefield at the rear edge of the acceleration field. It moves forward in the wakefield frame, and goes into the decelerating field. Dephasing in the range of decelerating field is the end of muon acceleration. However, the physics is quite different for the low-initial-energy muon acceleration in the beam-driven plasma wakefield. The muon's velocity cannot exceed that of the wakefield. The end of the muon acceleration is determined by the time of the consumed out of the electron-beam energy.

\begin{figure}[b]
\includegraphics[scale=0.1]{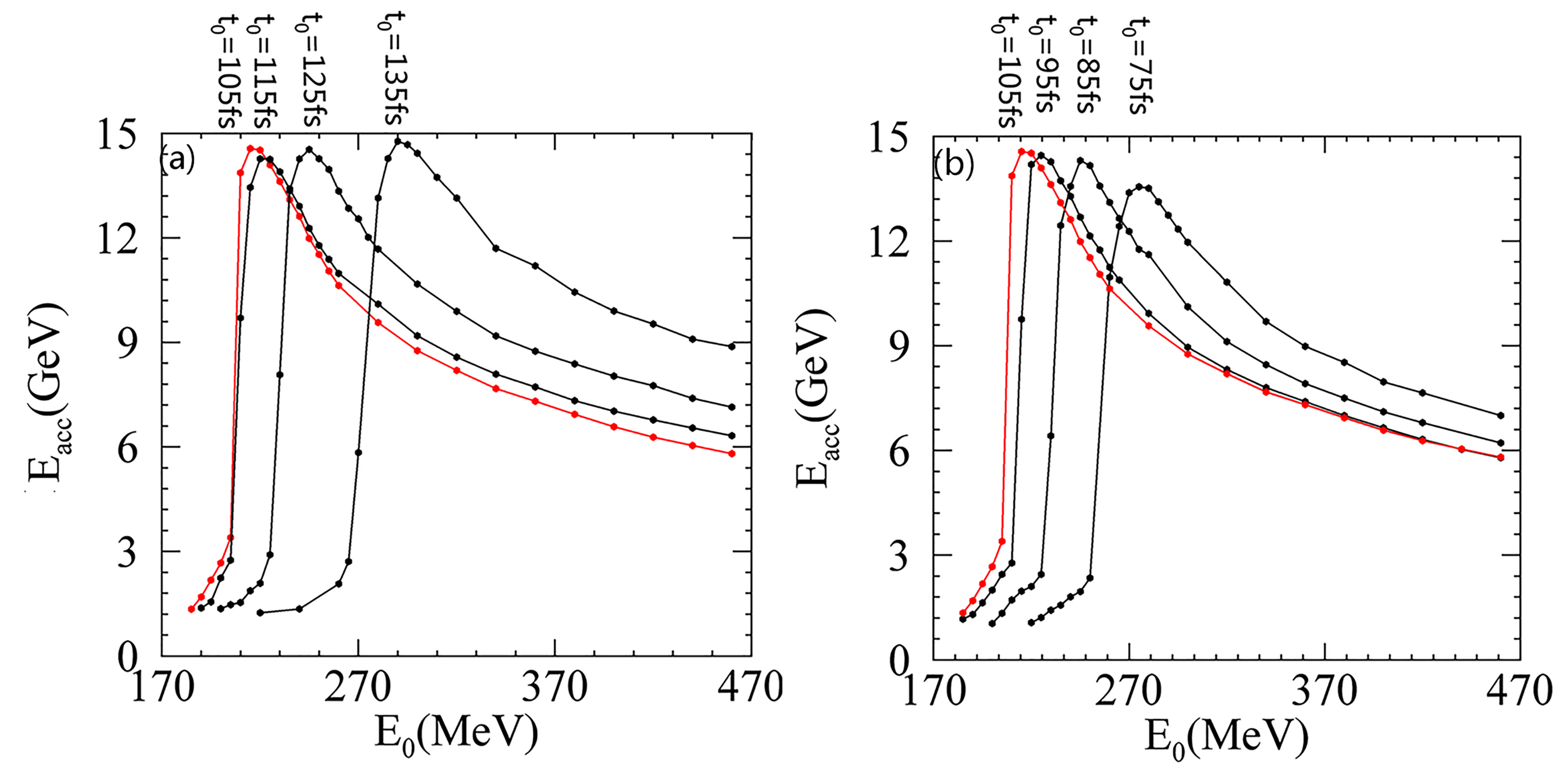}
\caption{\label{fig3}{\small\textbf{The relationship between the final acceleration energy and the initial energy of the muon beam for different injection time}}.{\small{ (a) and (b) shows the results when the initial position located in the acceleration filed and in the deceleration filed respectively. The red curves show the simulation results when the muon injection time $t_0=105\mathrm{fs}$. $E_{acc}$ stands for the final acceleration energy. The muons are located at the center line of the wakefield.}}}
\end{figure}

\textbf{The influence of the injection time and the initial energy of the muon acceleration process}. The relationship between the final acceleration energy and muon initial energy for different injection time is shown in Figure \ref{fig3} (a,b). The injection time determines the initial muon position. When the injection time is equal to $105\mathrm{fs}$, the initial muon position is located at the point b shown in Figure \ref{fig1}. Figure \ref{fig3} (a,b) show that there is a proper initial energy for every injection time to ensure that the final acceleration energy reaches the maximum, whether the initial position is in the acceleration region or in the deceleration region. This proper initial energy is defined as the critical initial energy, $E_{0c}$.

  For $E_{0}<E_{0c}$, the muon drops out of the wakefield from the rear side of the acceleration field before it breaks. When the injection time of muons is closer to $105\mathrm{fs}$, the final acceleration energy of muon increases with the initial energy as shown in Figure \ref{fig3} (a,b).

  a. The influence of the initial energy on the acceleration process: for the muons with the same injection time, the total acceleration time of muons increases with the initial energy. Therefore, the final acceleration energy of muons increases accordingly.

  b. The influence of the injection time on the acceleration process: for the muon with the same initial energy, when the muon initial position is closer to point b in the acceleration region, the final acceleration energy increases since the total acceleration time of muons in the acceleration filed increases. When the initial muon position is closer to the point b in the deceleration region, the energy loss decreases. As a result the final acceleration energy increases.

  For $E_0>E_{0c}$, the wakefield breaks before the muons reach the maximum acceleration field point in region A as shown in Figure \ref{fig1} .Therefore the total acceleration time of muons is equal to the existence time of wakefield. When the injection time of muons is close to $105\mathrm{fs}$, the final acceleration energy decreases with the increase of the initial energy as shown in Figure  \ref{fig3} (a,b).

  a. The influence of the initial energy on the acceleration process: for the muons with the same injection time, with the increase of the initial energy, muon can't reach the position where the acceleration field is larger in region A as shown in Figure \ref{fig1}. Therefore the final acceleration energy decreases.

  b. The influence of the injection time on the acceleration process: for the muons with the same initial energy, when the muon initial position is closer to point b in the acceleration region, the final acceleration energy decreases due to the smaller acceleration field. When the initial muon position is closer to point b in the deceleration region, the final acceleration energy decreases.  It is indicated that the muons are finally surpassed by the muons behind them at the beginning time.

  For $E_0=E_{0c}$, the total acceleration time of the muons stays in the acceleration region is always equal to the existence time of the wakefield. The corresponding initial critical energy, $E_{0c}$, reaches minimum for the injection time equal to $105\mathrm{fs}$, i.e., the optimal initial position at point b.

\begin{figure}[b]
\includegraphics[scale=0.05]{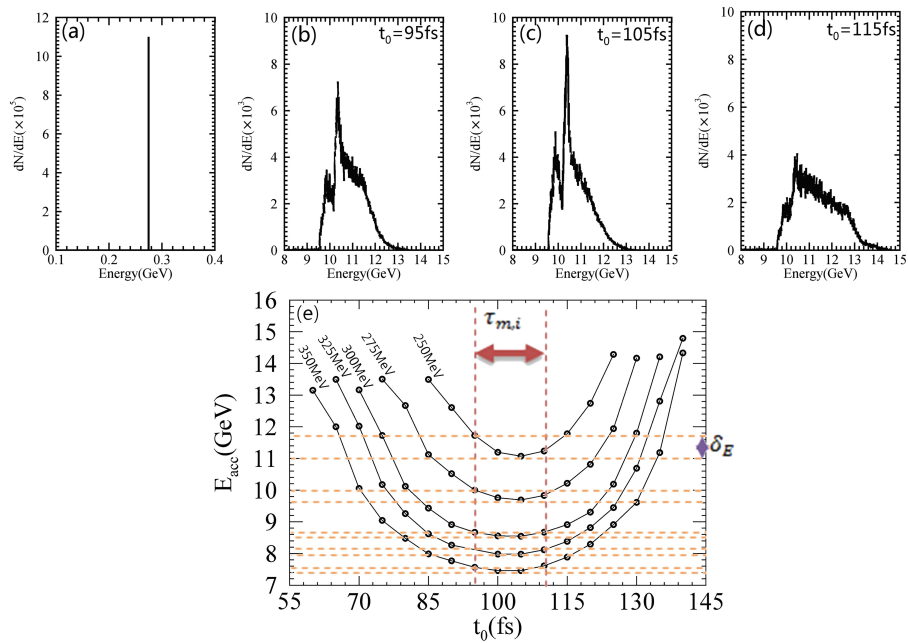}
\caption{\label{fig4} \small \textbf{The initial muon energy spectrum (a), the finally muon energy spectrum (b-d) and the final acceleration energy of muon versus the injection time for different initial energy (e)}. (a)  The Initial energy spectrum for an ideal mono-energetic muon beam with energy $275\mathrm{MeV}$. (b-d)  The finally energy spectrum for ideal mono-energetic muon beam for different injection time $t_0\in[95,105,115]$ and the simulation time $t=20.5\mathrm{ps}$. (e) The final accelerated energy of the muon beam. For the same pulse duration of the initial muon beam, $\tau_{m,i}$, the energy difference of the final muon beam, $\delta_E$, decreases with the initial energy.}
\end{figure}

 \textbf{The optimal injection time for energy spread.} When the plasma paramters and driven beam conditions are fixed, the final energy spread of the muon beam mainly depends on the injection time. Figure \ref{fig4} (b-d) show that the energy spread of the accelerated muon beam is narrowest for the injection time equal to $105\mathrm{fs}$. To explain this phenomenon, we analyze the relationship between the final acceleration energy and the injection time for different initial energy. With a serial of simulation results corresponding to a serial of initial mono-energetic muon beams locted one grid of the simulation box, the relationship between the final acceleration energy of muons and the injection time for different initial energy can be obtained shown in Figure \ref{fig4} (e). From Figure \ref{fig4} (e), for the same pulse duration of the initial muon beam, $\tau_{(m,i)}$, when the injection time is closer to $105\mathrm{fs}$, the energy difference of the final muon beam, $\delta_E$, decreases. As a result the muon beam energy spread will be narrower. This analyze is consistent with the simulation results shown in Figure \ref{fig4} (b-d). Furthermore, Figure \ref{fig4} (e) also shows that for the same pulse duration of the initial muon beam, $\tau_{(m,i)}$, the energy difference of the final muon beam, $\delta_E$, decreases with the initial energy. As a result, the energy spread of the muon beam is smaller for higher initial energy.

\begin{figure}[b]
\includegraphics[scale=0.15]{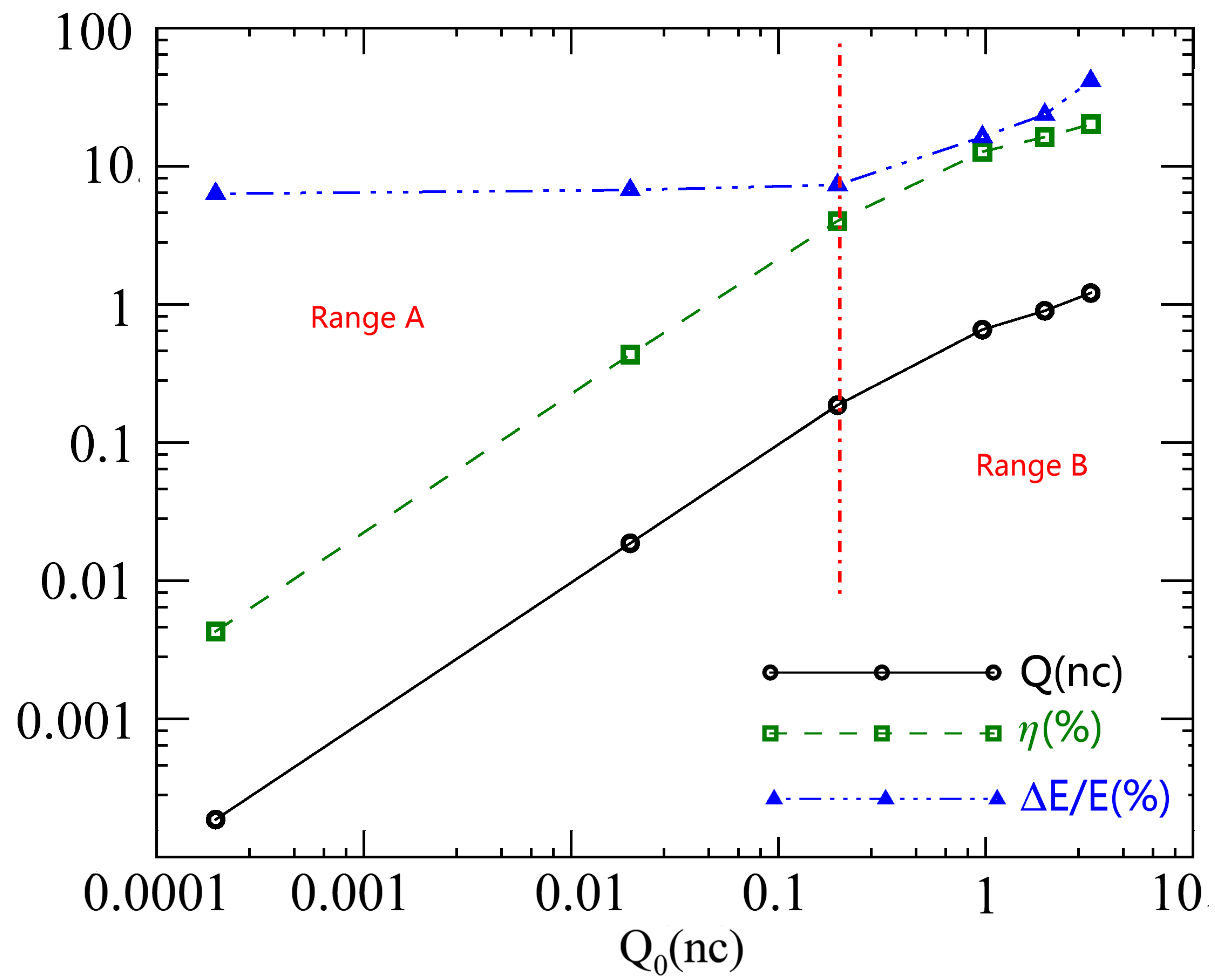}
\caption{\label{fig5} \small \textbf{The total charge Q, the energy transfer efficiency $\eta$ and the energy spread $\frac{\triangle E}{E}$ of the accelerated muon beam versus the initial total chargeof the muon beam,$Q_0$.} The blue trangle, green queral, black circle represents the simulation results of the total charge Q, the energy transfer efficiency $\eta$ and the energy spread $\frac{\triangle E}{E}$, respectively. The simulation parameters are same as those of Figure 2, besides the initial total charge $Q_0$. In Range A, $Q_0$ is less than 0.12nc, the efficiency, $\eta$, increases with $Q_0$. In Range B, the maximum efficiency can reach $20\%$ corresponding to $Q=1.2\mathrm{nC}$. }
\end{figure}

\textbf{The efficiency of the energy transfer from the driven electron beam to the muon beam.} Define the efficiency of the energy transfer from the driven electron beam to the muon beam as follows: $\eta = \frac{E_{\mu tot}}{E_{etot}}$, where the $E_{\mu tot}$ is the total energy of the accelerated muon beam and the $E_{etot}$ is the initial total energy of the driven electron beam. Figure \ref{fig5} shows that in Range A, the total charge Q of the accelerated muon beam is almost equal to the  initial total charge of the muon beam, $Q_0$, which means that little muons drop out of the wakefield during the acceleration process.  The energy spread $\frac{\triangle E}{E}$ is almost a constant about $6\%$. In Range B, for large $Q_0>1\mathrm{nC}$, the total charge of the accelerated muon beam has an upper limit, about $1.2\mathrm{nC}$. The energy transfer efficieincy can be reaches $20\%$ shown in Range B. In this case, the energy spread and the energy transfer are both increased with $Q_0$. However, the muon beam is still quasi-monoenergetic.

 In this study, it is proposed a new high-efficiency prompt acceleration scheme of a low-energy muon beam. The lifetime of the muons can be enlarged to several hundreds of microseconds within several tens of picoseconds. The longitudinal distribution of the low-energy muon beam is compressed due to the distribution of longitudinal wakefield $E_x$. The muon acceleration process is studied by three-dimensional PIC simulations. The simulation results show that the muon beam is accelerated from $275\mathrm{MeV}$ to $10\mathrm{GeV}$ within $22.05\mathrm{ps}$. The energy spread of the accelerated muon beam is about $6\%$. We have discussed the dependence of the final acceleration energy on the initial energy and the different injection time. For a given initial muon energy, an opitimal injection time exists to obtain the maximum muon energy and the minimun energy spread. In this prompt acceleration, a high energy transfer efficiency $20\%$ can be reached. Therefore, the total charge of the accelerated muon beam is about $1.2\mathrm{nC}$. In conclusion, with the high-efficiency prompt acceleration scheme, a compact high-energy moderate-flux muon source is coming and shows quite competitive compared with the exists low-energy muon beam from the traditional accelerator or the low-flux cosmic muons. In the near future, an energetic muon beam will be the most economical and hopeful way to bring the expected neutrino factory and the muon collider into reality and to find the lepton flavor violation and other new physics beyond the Standard Model.

   {\textbf{\large {\\Method}}}
   \textbf{\\The simulation parameters}. Our simulations are carried out using the three dimensional PIC code, EPOCH3d\textsuperscript{\cite{arber2015contemporary}}. The window:  $x\times y\times z=72\mathrm{\mu m}\times 60\mathrm{\mu m}\times 60\mathrm{\mu m}$ has velocity $v_w=2.9999\times10^8 \mathrm{m/s}$ which is very close to that of light in a vacuum. The number of grids is 240 and 120 in the longitudinal and transversal direction respectively. Electrons and protons fills the entire simulation box with a density  $n_p=1\times10^{19}\mathrm{cm^{-3}}$. An electron beam injects the simulation box along x direction at the beginning of the simulation and excites an extremely intense plasma wakefield. The electron beam is monoenegetic and the energy is $3\mathrm{GeV}$. The density distribution of the electron beam is the gauss distribution in three directions with $\sigma_x=10.6\mathrm{\mu m}$, $\sigma_y=\sigma_z=3.5\mathrm{\mu m}$. The number of electrons of the electron beam is $N=1\times10^{11}$ and the maximum number density is $n_{b,max}=7\times10^{25}m^{-3}$. There are two particles per cell for electrons, protons and four particles per cell for muons. Open conditions are applied for all boundaries. Different Figure has different muon beam parameters and different injection time. For Figure \ref{fig2}, the muon number of the muon beam is about $10^6$. The duration of the accelerated muon beam is compressed to $10\mathrm{fs}$ and the radius is focused to $5\mathrm{\mu m}$. The muon beam with initial energy of $275\mathrm{MeV}$ is injected in the simulation window at $105fs$ after the electron beam is injected. For Figure \ref{fig3} and Figure \ref{fig4} (e), the muons fill in one grid in the simulation box and locate at the center line of the three-dimensional wakefield.  The maximum density is unchanged. For the Figure \ref{fig3}, the initial muon energy ranges from $180\mathrm{MeV}$ to $460\mathrm{MeV}$, and the muon injection time ranges from $75\mathrm{fs}$ to $135 \mathrm{fs}$. For  Figure \ref{fig4} (e), the initial muon energy ranges from $250\mathrm{MeV}$ to $350\mathrm{MeV}$, and the muon injection time ranges from $60\mathrm{fs}$ to $135\mathrm{fs}$. For Figure \ref{fig4} (b-d), only the injection time is different from Figure\ref{fig1} and ranges from $95\mathrm{fs}$ to $115\mathrm{fs}$. For Figure \ref{fig5}, only the initial charge of the muon beam is different from Figure\ref{fig1} and ranges from $0.0002nc$ to $3.2nc$.

\begin{acknowledgments}
\end{acknowledgments}

 {\textbf{\large {\\Data availability}}}. The data that support the findings of this study are available
from the corresponding authors on request.

  \nocite{*}




\end{document}